\documentclass[prl,twocolumn,showpacs]{revtex4}
\usepackage{graphicx}
\usepackage{amssymb}

\begin{document}


\title{From thermally activated to viscosity controlled fracture of
  biopolymer hydrogels }

\author{T. Baumberger and O. Ronsin}
\affiliation{INSP, UPMC Univ Paris 06, CNRS UMR 7588
140 rue de Lourmel, 75015 Paris France}

\date{\today}

\begin{abstract}

We report on rate-dependent fracture energy measurements over three decades of 
steady crack velocities in alginate 
and gelatin hydrogels.  
We evidence that, irrespective of gel thermo-reversibility,  thermally activated
``unzipping" of the non-covalent cross-link zones results in slow 
crack propagation, prevaling against the toughening effect of viscous solvent drag
during chain pull-out, which becomes efficient above a few mm.s$^{-1}$.  
We extend a previous model [Baumberger {\it et al.} Nature
Materials, {\bf 5}, 552 (2006)] to account for both mechanisms, 
 and estimate the microscopic unzipping rates.

\end{abstract}


\maketitle

A variety of biopolymers (e.g. proteins and polysaccharides)
can self-assemble into
physically cross-linked networks in aqueous solutions. The resulting
hydrogels are usually
bio-compatible and show mechanical properties that mimic
those of extracellular matrices. They are therefore good candidates
as scaffolds for {\it in vivo} tissue regeneration, e.g. cartilage 
\cite{Tissue}.
Such applications require sufficient mechanical
strength to withstand manipulations associated with
implantation and in vivo existence \cite{Mooney}.
Surprizingly, understanding the physical  mechanisms at work during 
the ultimate
behavior (yield and failure) of
non-covalently cross-linked hydrogels is still in its infancy.
In this context, Baumberger {\it et al.} \cite{Nous} have
identified
a basic dissipative process
in the slow fracture of
gelatin which they accounted for in a minimal model: gelatin gels
belong to a wider class of biopolymer
networks with  extended cross-link zones, e.g. H-bond stabilized
multiple helices, distributed along the chains and acting as mechanical
fuses which may unzip or unreel under
tension \cite{Reel}, forcing
{\it overall} polymer
pull-out without chain scission, at the cost of the viscous dissipation due to the solvent
drag on the whole chain.
This highly efficient mechanism makes these nominally weak gels
amazingly resistant to crack propagation  even at moderate 
velocities of $1$  mm.s$^{-1}$ and above. In contrast, under quasi-static
loading physical gels generally creep under stress \cite{Higgs} and ultimately
break after a strongly stress-dependent, random delay  \cite{Bonn}.
Though this low rate behaviour, especially  in thermoreversible
networks, undoubtly pinpoints the role of thermal activation,
it remains unclear whether the rate-limiting process in fracture of
such
intrinsically
disordered materials is the nucleation of
unstable cracks \cite{Bonn} or their subsequent growth\cite{Bonncomment}.
Generally speaking, unraveling nucleation and growth effects  is  a
subtle task \cite{Vanel} and, when feasible experimentally,
studying low velocities, {\it steady state} crack propagation  provides
more straightforward insight into the fracture mechanisms.
Dealing with  gelatin gels, however, reliable
measurement of both small crack-tip velocities and the associated 
energy release
rates is hindered by two intrinsic limitations:
(i) Thermal activation promotes crosslink rearrangements in the
bulk as well, hence stress relaxation which competes with crack
propagation to release elastic energy;
(ii) Strong pinning of the crack front by network inhomogeneities
may result in front instabilities \cite{Morpho} which makes  tip
position and fracture area measurements ambiguous \cite{Tanaka}.
As reported in this letter we circumvent these drawbacks by using
alginate gels in which
cross-linking is ensured by ionic bridges along extended zones between
polyelectrolyte chains, with
binding energies intermediate between H-bond and covalent ones.
Though {\it non}-thermoreversible, these gels can be termed physical since
interchain bounds remain weaker than backbone ones.
Taking advantage of their negligible creep under small stresses
\cite{Shull} and of the absence of
crack-front instability down to
velocities in the $10\,\mu$m.s$^{-1}$ range, we clearly evidence that at low
enough rates, crack propagation in these gels is thermally activated; moreover,
after checking that they also exhibit a viscosity controlled
regime, we extend a previous model \cite{Nous} to account for
thermally activated, stress-aided cross-link dezipping.
The outcome is an estimate of the debonding rate which lends further
support to the prevailing picture of ``egg-box" binding zones in
alginate \cite{Perez}.

\begin{figure}[h]
     \includegraphics{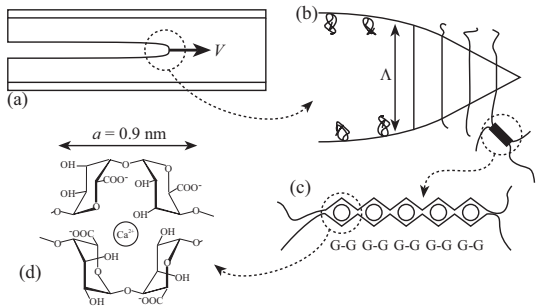}
     \caption{Schematic  hierachy of length
     scales in the fracture of an alginate gel. (a) Experimental
     set-up showing the mm-scale crack opening (b) Crack-tip region.
      (c) Egg-box representation of a  ionic
     cross-link. (d) Molecular picture of a binding  unit.}
     \label{fig:setup}
\end{figure}

Alginates are
polysaccharides
composed of  sequences of two
sugar residues, referred to as G and M \cite{Polysacch}. According to
the ``egg-box"
model \cite{Perez}, gelation occurs via
interchain chelation of divalent ions (eggs), here Ca$^{2+}$, at specific
sites (boxes) formed by the assembly of two subsequences made of
several
contiguous G-units (Fig. \ref{fig:setup}).
The length of a binding GG pair is $a = 0.9$ nm.

Sodium alginate was purchased from Kalys
(France). From supplier's specifications (average molecular weight
216 kDa and 55\% ``M" residues) the average contour
length is $\Lambda = 550$ nm. We use different alginate 
concentrations $c$ between
$0.5$ and $2.5$ g for $100$ ml of solvent (deionized water, otherwise
specified) and a constant ratio of $10^{-3}$ mol of Ca$^{2+}$ for
$1$ g of alginate known to minimize gel shrinking  (syneresis).
Homogeous gel samples are obtained \cite{CaCO3} by in 
situ progressive release of
Ca$^{2+}$ ions from insoluble CaCO$_{3}$ particles subsequent to the
addition of a slowly hydrolyzing acid (``GDL" from Sigma) quickly
mixed with the pregel solution before casting it in a mold.
The samples --- $30$ cm long slabs, $1$ cm thick and $3$ cm wide,
attached to parallel grips --- are stretched along
their widths. Details of the set-up have been given previously
\cite{Nous}. All experiments are performed at $T = 295$~K. Cracks
are initiated by notching in the mid-plane.
When  micron-sized CaCO$_{3}$ particles (from Sigma) are used, 
gelling is slow and
homogeneous. However, at low $c$,  particle sedimentation occurs
leading to a toughness gradient as revealed by  a tilted  crack front.
With nanoparticles of average size $90$ nm (American
Elements, CA), no sedimentation occurs but gelation is much faster
and, at high $c$, yields inhomogeneous samples as revealed by a non-planar,
tortuous crack front. We have therefore used a mixture of nano- and
micro- particles, in such proportions as to yield straight crack
fronts perpendicular to the slab faces and propagating along its mid-plane,
a very stringent requirement.

Steady crack velocities $V$ are measured by video tracking of the crack
tip. In our fixed grip configuration, the energy release rate $\mathcal G$ is imposed.
We compute it, neglecting edge effects,
from the total stored elastic
energy as determined from the force {\it vs.} stretching ratio
loading curve
measured on an unnotched sample.
The small strain shear
modulus  $\mu = 1.5$~kPa for $c = 1.5\%$ 
varies approximately as $c^{2}$ in the studied range.

\begin{figure}[h]
     \includegraphics{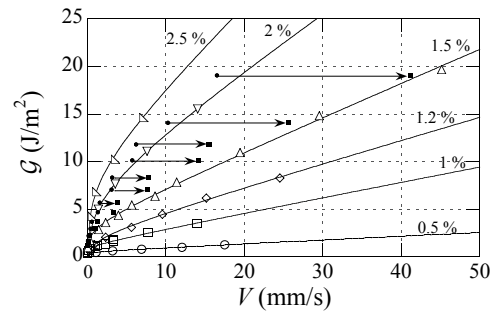}
     \caption{Fracture energy vs. tip velocity for different
     alginate/water wt. fraction (open symbols). Filled circles are
     for a $c=1.5$ \% gel in water/glycerol. Filled
     squares: same data set as a function of velocity scaled by
     solvent viscosity ratio (see text). Lines are best fits by eq. 
\ref{eq:Gmodel}.  }
     \label{fig:allc}
\end{figure}

Figure \ref{fig:allc} displays $\mathcal G(V)$ curves for alginate gels with
different  $c$.  They share with gelatin ones the principal features 
we associate with a visco-plastic fracture mechanism via 
scissionless  chain 
pull-out \cite{Nous}: 
a quasi-linear strong growth at high $V$, extrapolating at $V=0$ to a non-zero
threshold value of order a few J.m$^{-2}$, with a slope increasing 
linearly with the solvent viscosity $\eta$. This latter property is 
confirmed in a control experiment:  using a glycerol/water mixture we 
induce a $2.5$-fold increase of $\eta$, resulting is a clear 
toughening with respect to a pure water-based gel.  When plotted vs. 
$\eta V$, however, both $\mathcal G$ curves
collapse over the whole
$V$-range  (Fig.\ref{fig:allc}).

Beyond these remarkable similarities (irrespective of the 
thermoreversibility of the gels), $\mathcal G(V)$  curves for 
alginate specifically
exhibit a systematic round-off
at lower velocities, which implies no clear threshold for 
crack propagation. This is reasonably ascribable to thermally 
activated unzipping. 
Since with these systems $T$ cannot be
varied widely enough to build a significant Arrhenius plot, we must 
rely on modeling to
test this hypothesis.
This leads us to  extend our previous model \cite{Nous} to account for 
thermal activation.
In this description, the complex network
features are lumped into a few parameters, namely the average
chain contour length $\Lambda$, the areal density $\Sigma_{0}$ of 
chains crossing
the fracture plane, the size $a$ of a binding
unit and the activation barrier $U$.
The fracture energy is then computed according to the
Dugdale-Barenblatt theory \cite{Maugis} assuming a
uniform stress $\sigma_{tip}$ over a small-scale cohesive zone at the crack
tip. The fracture criterion corresponds to the overall
pull-out of chains, stretched taut, i.e. to a maximum  opening $\Lambda$ of
the tip in the cohesive zone (Fig. \ref{fig:setup}b). The fracture energy is thus simply:
\begin{equation}
     \mathcal G = \sigma_{tip}\Lambda
     \label{eq:Dugdale}
\end{equation}

Note that $\mathcal G\gtrsim 1$ J.m$^{-2}$ entails $\sigma_{tip}
\gtrsim 2$ MPa $\gg \mu$. This justifies that the chains are almost
fully stretched in the crack tip vicinity. Accordingly,
if $\nu$ is the frequency at which units are
released, neglecting re-bonding rate, the pull-out velocity reads:
$\vartheta = a\nu$.

There is
little hope to compute the very shape of the cohesive pre-crack  due to the
strongly non-linear, anisotropic elastic field which prevails ahead of the
tip, where stresses reach values of order $\sigma_{tip}\gg\mu$
\cite{Hui}. Instead we assume a wedge-shaped tip, which provides
a simple kinetic relationship~:
\begin{equation}
     \vartheta = a\nu = \alpha V
     \label{eq:cine}
\end{equation}
where  the fitting parameter $\alpha$ should consistently be $\gg 1$ 
to account for crack blunting \cite{Hui}.

Closure of the problem requires relating $\sigma_{tip}$ to $\vartheta$ hence to
$\nu$. The chain tension
decreases along the polymer away from the crack edge due to a viscous drag of order
$\eta\vartheta$ per unit of contour length. Let us make a crude
estimate of the tension $f_{Y}$ under which reels yield as:
\begin{equation}
     f_{Y} \simeq \frac{\sigma_{tip}-p}{\Sigma_{0}}-\eta\Lambda\vartheta
     \label{eq:tension}
\end{equation}
The stress $p$ stems from the capillary pressure jump which tends to
suck-in the chain when it is pulled-out dry. It is zero when the
crack tip opening is wetted by a drop of solvent \cite{Nous} but is $p=
\epsilon_{H}\Sigma_{0}/a$ for a dry  tip, with $\epsilon_{H}$ the
free energy of solvation per residue of length $a$ (the work done by
the capillary force against drawing a residue into the gap).

Following Kramers \cite{Hanggi}, the bond breaking rate in the biased 
binding potential is given
by:
\begin{equation}
     \nu = \nu_{0}\exp[-(U-f_{Y}a)/k_{B}T]
     \label{eq:Kramers}
\end{equation}
where $\nu_{0}$ is the frequency of attempt to escape over the
binding barrier.

Eqs. (\ref{eq:Dugdale})--(\ref{eq:Kramers}) yield~:
\begin{equation}
     \mathcal G = \Delta\mathcal G_{H}+ \mathcal
G_{0}\left[1+\frac{k_{B}T}{U}\ln(V/V^\star)+  \gamma\eta
V\right]
     \label{eq:Gmodel}
\end{equation}
with
\begin{equation}
     \mathcal G_{0} = \frac{U\Lambda\Sigma_{0}}{a},\,
     \Delta\mathcal G_{H} = \frac{\epsilon_{H}\Lambda\Sigma_{0}}{a},\,
V^\star = \frac{a\nu_{0}}{\alpha},\,\gamma = \frac{a\Lambda\alpha}{U}
     \label{eq:par}
\end{equation}

As seen on Fig. \ref{fig:allc},
the functional form of $\mathcal G(V)$ in eq.(\ref{eq:Gmodel}) 
provides an excellent overall fit of
experimental data for gels of composition $c$ ranging from $0.5$
\% to $2.5$ \%. In order to put this analysis on a more quantitative footing,
we note that the
expression of
eq. (\ref{eq:Gmodel}) features three independent fitting parameters
only. Disregarding  multiplicative constants of order unity in eq.
(\ref{eq:par}), and setting $\Lambda$ and $a$
to their nominal average values, there remain four parameters actually
unknown~:  $U$, $\alpha$, $V^\star$ and $\mathcal G_{0}$. 
$\Delta\mathcal G_{H}$ is
determined through an independent experiment where the crack tip is
wetted by a drop of the solvent~; for gelatin gels \cite{Nous}, this 
results in a
shifted $\mathcal G(V)$ curve. For alginate gels, we have observed no significant effect of
wetting the tip with water (with NaCl added to equilibrate the
Na$^{+}$ concentration with that of the bulk sodium alginate one).
We conclude that the polyelectrolyte alginate
chains are pull-out in a hydrated state, hence $\Delta\mathcal
G_{H}\simeq 0$.

\begin{figure}[h]
     \includegraphics{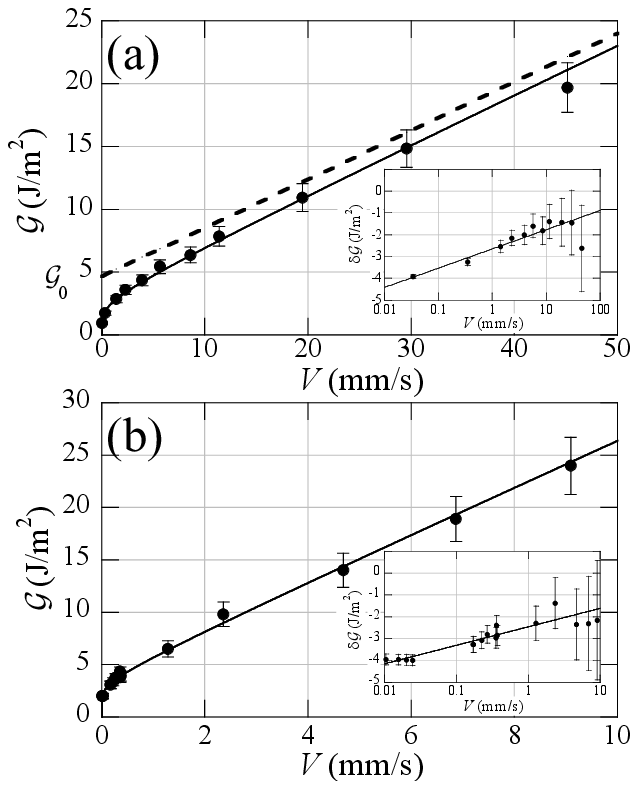}
     \caption{Fit (solid curve) of $\mathcal G(V)$ data using
     eq.(\ref{eq:Gmodel}) for (a) $c = 1.5$ \% alginate and
     (b) $c=10$ \% gelatin. 
     Insets : Residual energy $\delta \mathcal G$ after subtraction of the linear
     component. Dashed line on (a) is the $T=0$ K characteristics. }
     \label{fig:Fit}
\end{figure}

The relatively simple eggbox structure of interchain binding in alginate
prompts
us to choose the activation barrier height $U$ as an input for data
fitting with eq.(\ref{eq:Gmodel}). To do so, we compute the electrostatic
energy of
a Ca$^{2+}$ ion assuming it is involved in
purely ionic bonds with its nearest neighbours  O atoms in the GG$<>$GG 
cage (Fig. \ref{fig:setup}d) With an average
Ca--O distance $r=2.3 \AA$ \cite{Perez}, this entails
$U/ k_{B}T = 4l_{B}/r = 12$ at  $295$ K in water where
the Bjerrum length $l_{B} $ is $ 7 \AA$.

We now restrict our analysis to a $c=1.5$~\% gel for which we have got the
most extensive data set although the same qualitative conclusions can be
drawn from the other concentrations. The corresponding fit  is shown on
Fig. \ref{fig:Fit}a. The Griffith, or ``$T=0$ K" energy threshold  is 
$\mathcal G_{0}
\simeq 4.6 $ J.m$^{-2}$. A mere
extrapolation to $V = 0$ of the apparent linear regime over the experimental
velocity window underestimates this value. Strictly speaking, the purely
viscous regime is only reached asymptotically for $V\to V^\star$. Here,  $V^\star \simeq
0.6$ m.s$^{-1}$, out of experimental range.  One can however check
on eq.(\ref{eq:Gmodel})
that the correction to the {\it slope} of $\mathcal G(V)$ due to the
activated term is negligible for $V$ larger than $1$ mm.s$^{-1}$,
i.e. that the asymptotic slope can be safely estimated from the low
velocity ($V < V^\star$) data set. The value $\alpha \simeq 8$ obtained
from this slope is indicative of a strongly blunted tip
\cite{Nous,Hui}.

Although the fitting value for $V^\star$ is strongly sensitive to
approximations in the trial value for $U$,  we claim that the
order of magnitude of the attempt frequency  $\nu_{0} \simeq 5.5\times
10^{9}$ s$^{-1}$  yields deep insight into the unzipping dynamics.
As
discussed by Evans \cite{Evans}, in liquids,
the thermal impulses that drive unbinding events are damped
by viscous coupling to the environment. Accordingly, the opening of a
GG$<>$GG
molecular cage
is expected to induce hydrodynamic flow over a size of order $a$ hence a
damping rate scaling as $k_{B}T/\eta a^{3} = 5.4\times
10^{9}$ s$^{-1}$ in water at $295$ K. This is precisely the order
of magnitude of $\nu_{0}$, lending strong support to a simple
unzipping scenario where calcium ions would be released one by one,
as might be schematically expected from the egg box picture of the
cross-links. This is, to our best knowledge, the first time a {\it 
dynamical} argument is given in favor of the egg-box architecture. 

On approaching $V^{\star}$,  the 
activation
barrier smoothes out and the escape rate is no longer
given by eq.(\ref{eq:Kramers}) since advection-driven, deterministic
debonding events becomes increasingly prevalent. 
In the opposite range, as $V\to
0$, re-binding events must become relevant and lead to a
 regime ruled by the slow creep of the cross-links themselves, with
$f_{Y}$ going linearly to zero \cite{Higgs}. Consequently,  
eq.(\ref{eq:Kramers}) becomes unphysical for $f_{Y}\to 0$
i.e.  for $V \lesssim V_{min} =
V^\star \exp(-U/k_{B}T)$. For alginate gels, we estimate $V_{min}
\simeq 4\,\mu$m.s$^{-1}$, well below the lower bound of our velocity
set.  This  legitimates {\it a posteriori}  using eq.(\ref{eq:Gmodel})
over the whole experimental window, which fulfills the requirement $V_{min}\ll 
V\ll V^\star$.  

The previous discussion is based on generic features of zipper-like 
cross-linked gels. It can be therefore expected that thermally 
activated, stress-aided dezipping will be all the more efficient  
in thermoreversible gels.
This prompts us to 
reassess, at least qualitatively,  the case of gelatin. As already mentioned, this requires taking special care.
First of all, using a $c = 10$ \%
gelatin/water gel, the crack front instability \cite{Morpho} is 
pushed down below $V_{c} =
20\,\mu$m.s$^{-1}$. Moreover, we set-up a procedure  to correct the 
fracture energy for
stress relaxation. For this purpose, we   record
the crack velocity {\it vs.} time in response to various crack
openings. A twin sample, kept unnotched, is submitted to the very 
same stretching sequence
while recording the loading force. The stored elastic energy is
computed assuming that stress relaxation in such weak, transient
networks occurs via debonding of cross-links which eventually rebind
at a more favourable place, therefore resulting in a drift of the
reference state, hence in a mere shift of
the non-linear stress-strain curve \cite{Elias}. As shown on fig.\ref{fig:Fit},
this tedious procedure ultimately  reveals a
clear logarithmic behaviour over at least two decades,
essentially below $1$ mm.s$^{-1}$, which 
was therefore overlooked in previous studies \cite{Nous}.  

According to the present model, breaking one-by-one H-bonds ($U\simeq 0.1$ eV)
between peptidic
residues distant of $a = 0.3$ 
nm would result in a logarithmic shift of the fracture energy of 
water-based gels 
strictly between $V^{\star} =  4$ m.s$^{-1}$ and 
$V_{min} = V^{\star} \exp(-U/k_{B}T) = 8$ cm.s$^{-1}$ where we use  
the conservative value $\alpha = 10$ for the blunting parameter \cite{Nous}. 
Both the span of this velocity bracket and its absolute location are 
clearly incompatible with the experimental observation.
We are therefore led to
propose that the basic thermal event involves the {\it cooperative}
debonding of $n$ several subsequent H-bonds along a
cross-link zone. This we attribute to the strong topological
constraint imposed by the triple-helix structure of the cross-links
\cite{Nij},
the unzipping of which requires also large-scale unwinding. This
cooperative mechanism
cannot be discriminated from a one-by-one unzipping as long as one
probes (through $\mathcal G_{0}$) the yield tension $f_{Y}$ which reads $nU/na=U/a$. However,
the characteristic unbinding rates  are dramatically affected by the 
effective barrier energy ($\sim nU$) and by
the bulkiness of the activated
unit ($\sim na$). On this respect, a cooperativity level 
of $n\approx 3$ would  make gelatin gels looking similar to alginate 
ones, everything equal otherwise. 

This analysis confirms the importance of thermally
activated rate processes in soft matter fracture physics
\cite{Vanel,Chaud}. Our claim that the ``subcritical'' fracture behaviour is sensitive to distinctive 
topological features of
zipper-like cross-linked networks opens the way to a more extensive
experimental study, taking advantage of the wealth of network
architecture offered by biopolymer hydrogels.

\begin{acknowledgments}
We are indebted to C. Caroli for in-depth comments and careful
reading of the manuscript.
\end{acknowledgments}

\end{document}